\documentclass[twocolumn]{article}
\usepackage[utf8]{inputenc}
\usepackage{textcomp,marvosym}
\usepackage[table,xcdraw]{xcolor}
\usepackage{sectsty}
\allsectionsfont{\sffamily}
\usepackage{abstract}

\usepackage{arydshln}
\usepackage{url}
\usepackage{listings}
\usepackage{hyperref}
\usepackage{float}
\usepackage{enumitem}
\setlist[enumerate]{noitemsep}
\usepackage{nameref,hyperref}
\usepackage{textgreek}
\usepackage{graphicx}
\usepackage{multirow}

\usepackage[left=0.6in,right=0.6in,top=0.6in,bottom=1in]{geometry}
\usepackage{authblk}
\usepackage[backend=biber,style=nature,sorting=none]{biblatex}
\providecommand{\keywords}[1]{\textbf{\textit{Keywords --}} #1}
\usepackage[font=small]{caption}

\begin{document}

\begin{refsection}

\title{\sffamily Extremist ideology as a complex contagion: the spread of far-right radicalization in the United States between 2005-2017\rmfamily}

\date{}

\author[a,b,1]{\normalsize Mason Youngblood}

\affil[a]{\scriptsize Department of Psychology, The Graduate Center, City University of New York, New York, NY, USA}
\affil[b]{\scriptsize Department of Biology, Queens College, City University of New York, Flushing, NY, USA\newline\textsuperscript{1}myoungblood@gradcenter.cuny.edu}

\twocolumn[
\begin{@twocolumnfalse}
	
\maketitle
	
\vspace*{-20pt}
\begin{abstract}
\vspace*{-10pt}
		
Increasing levels of far-right extremist violence have generated public concern about the spread of radicalization in the United States. Previous research suggests that radicalized individuals are destabilized by various environmental (or endemic) factors, exposed to extremist ideology, and subsequently reinforced by members of their community. As such, the spread of radicalization may proceed through a social contagion process, in which extremist ideologies behave like complex contagions that require multiple exposures for adoption. In this study, I applied an epidemiological method called two-component spatio-temporal intensity modeling to data from 416 far-right extremists exposed in the United States between 2005 and 2017. The results indicate that patterns of far-right radicalization in the United States are consistent with a complex contagion process, in which reinforcement is required for transmission. Both social media usage and group membership enhance the spread of extremist ideology, suggesting that online and physical organizing remain primary recruitment tools of the far-right movement. Additionally, I identified several endemic factors, such as poverty, that increase the probability of radicalization in particular regions. Future research should investigate how specific interventions, such as online counter-narratives to battle propaganda, may be effectively implemented to mitigate the spread of far-right extremism in the United States.

\\\\
		
\end{abstract}
\end{@twocolumnfalse}]

\section*{Introduction}

The far-right movement, which includes white supremacists, neo-Nazis, and sovereign citizens, is the oldest and most deadly form of domestic extremism in the United States \cite{Simi2017,Piazza2017}. Despite some ideological diversity, members of the far-right often advocate for the use of violence to bring about an ``idealized future favoring a particular group, whether this group identity is racial, pseudo-national, or characterized by individualistic traits'' \cite{Pirus2017}. Over the last decade, the far-right movement was responsible for 73.3\% of all extremist murders in the United States. In 2018, this statistic rose to 98\% \cite{TheAnti-DefamationLeagueCenteronExtremism2019}. The increasing severity of far-right extremist violence, as well as the associated rhetoric on social media \cite{Winter2019,Davey2019}, has generated public concern about the spread of radicalization in the United States. Former extremists have referred to it as a public health issue \cite{Allam2019,Bonn2019}, an idea advocated for by some policy experts as well \cite{Weine2016,Sanir2017}.

There is little evidence that radicalization is primarily driven by psychopathology \cite{Post2015,Webber2017,Misiak2019}. Rather, radicalization appears to be a process in which individuals are destabilized by various environmental factors, exposed to extremist ideology, and subsequently reinforced by members of their community \cite{Webber2017,Jasko2017,Jensen2018,Mills2019,Becker2019}. Even ``lone wolves'', or solo actors, often interact with extremist communities online \cite{Kaplan2014,Post2015,Holt2019}. As such, radicalization may spread through a social contagion process, in which extremist ideologies behave like complex contagions that require multiple exposures for adoption \cite{Guilbeault2018}, which has been observed for political movements more broadly \cite{Gonzalez-Bailon2011}. Previous research suggests that extremist propaganda \cite{Ferrara2017}, hate crimes \cite{Braun2010,Braun2011}, intergroup conflict \cite{Buhaug2008,Gelfand2012}, and terrorism \cite{Midlarsky1980,Cherif2009,LaFree2012,White2016a,White2016b} exhibit similar dynamics.

The environmental factors favoring radicalization, referred to here as endemic factors, include variables like poverty rate that may influence individuals’ risk of adoption in particular regions. As such, endemic factors have the potential to enhance or constrain the spread of contagions through populations across geographic space. Although significant research has been done on how endemic factors predict radicalization (and resulting violence) \cite{McVeigh2004,Goetz2012,LaFree2014,Piazza2017,Medina2018}, few studies have investigated how these factors interact with contagion processes. The aim of the current study is to determine whether patterns of far-right radicalization in the United States are consistent with a contagion process, and to assess the influence of critical endemic factors. After controlling for population density, I assessed the following endemic factors that have been implicated in previous research on radicalization, extremism, and mass shootings: poverty rate \cite{Gale2002,Durso2013,Suttmoeller2015,Suttmoeller2016,Piazza2017,Lin2018,Suttmoeller2018,Medina2018,Kwon2019a}, unemployment rate \cite{Green1998,Jefferson1999,Gale2002,Espiritu2004,Goetz2012,Piazza2017,Majumder2017,Pah2017}, income inequality \cite{McVeigh2004,Goetz2012,McVeigh2012,Majumder2017,Kwon2017,Kwon2019a,Kwon2019b}, education levels \cite{Espiritu2004,Florida2011,Durso2013,McVeigh2014,Gladfelter2017,Kwon2017,Kwon2019b}, non-white population size \cite{McVeigh2004,LaFree2014,Gladfelter2017,Medina2018}, violent crime rate \cite{McVeigh2012,Gladfelter2017,Sweeney2018}, gun ownership \cite{Pah2017,Anisin2018,Lin2018,Reeping2019}, hate groups per capita \cite{Adamczyk2014}, and Republican voting \cite{McVeigh2014,Medina2018}. Furthermore, I aim to determine whether individual-level variables, such as social media use, enhance the spread of far-right radicalization over space and time. Social media platforms increasingly appear to play a role in radicalization, both as formal recruitment tools \cite{Wu2015,Bertram2016,Aly2017,Awan2017} and spaces for extremist communities to interact \cite{Amble2012,Dean2012,Winter2019,Pauwels2014}. If social media platforms augment physical organizing \cite{Bowman-Grieve2009,Holt2016}, then they may also enhance the spread of radicalization.

Although social media platforms relax geographic constraints on communication, evidence suggests that social media networks still exhibit spatial clustering. For example, the majority of an individuals’ Facebook friends live within 100 miles of them \cite{Bailey2018}, the probability of information diffusion on social media decays with increasing distance \cite{Liu2018}, and online echo chambers map onto particular locations \cite{Bastos2018}. Since complex contagions require reinforcement, and the majority of online friendship ties are within a close radius, the diffusion of extremist ideologies online should still exhibit some level of geographic bias.

In order to model the spread of far-right radicalization I used a two-component spatio-temporal intensity (twinstim) model \cite{Meyer2017}, an epidemiological method that treats events in space and time as resulting from self-exciting point processes \cite{Reinhart2018}. In this framework, future events depend on the history of past events within a certain geographic range. Event probabilities are determined by a conditional intensity function, which is separated into endemic and epidemic components. This allows researchers to assess the combined effects of both spatio-temporal covariates and epidemic predictors. Epidemic, in this framework, refers to any level of contagion effect and does not necessarily imply uncontrollable spread. With a couple of notable exceptions \cite{Zammit-Mangion2012,Clark2018}, previous applications of self-exciting point process models in terrorism and mass shooting research have not simultaneously modeled diffusion over both time and space \cite{Porter2010,Lewis2012,White2013,Towers2015,Garcia-Bernardo2015,Tench2016,Johnson2017,Collins2020}.

The radicalization events in this study, which correspond to where and when a radicalized individual's extremist activity or plot was exposed, came from the Profiles of Individual Radicalization in the United States (PIRUS), an anonymized database compiled by the National Consortium for the Study of Terrorism and Responses to Terrorism (START) \cite{Pirus2017}. PIRUS is compiled from sources in the public record, and only includes individuals radicalized in the United States who were either arrested, indicted, or killed as a result of ideologically-motivated crimes, or were directly associated with a violent extremist organization. I chose to use PIRUS instead of the Terrorism and Extremist Violence in the United States (TEVUS) database because events in PIRUS are disambiguated by individual and include social variables that may influence the diffusion process.

A contagion effect in this modeling framework could result from one of two forces. The first is a copy-cat effect, in which individuals copy behaviors observed directly or in the media. Although this effect has been proposed in terrorism and mass shooting research in the past \cite{Nacos2009,Towers2015}, it seems to be a more plausible contagion mechanism for specific methods of violence \cite{Helfgott2015} (e.g. suicide bombings \cite{Tominaga2018}) rather than radicalization more broadly. The second is linkage triggered by activism and organizing, or ideologically-charged events (e.g. elections, demonstrations, policies), in that region. To differentiate between these two forces, I included two sets of epidemic predictors in the modeling. The first two event-level variables, plot success and anticipated fatalities, might be expected to increase epidemic probability if a copy-cat effect is present. This is because successful large-scale events are probably more contagious due to increased media coverage \cite{Towers2015}. Alternatively, the two individual-level variables, group membership and social media use, might be expected to increase epidemic probability if activism and organizing drive the linkage between events.

\section*{Methods}

\subsection{Data Collection}

\begin{figure}
\centering
\includegraphics[width=0.9\linewidth]{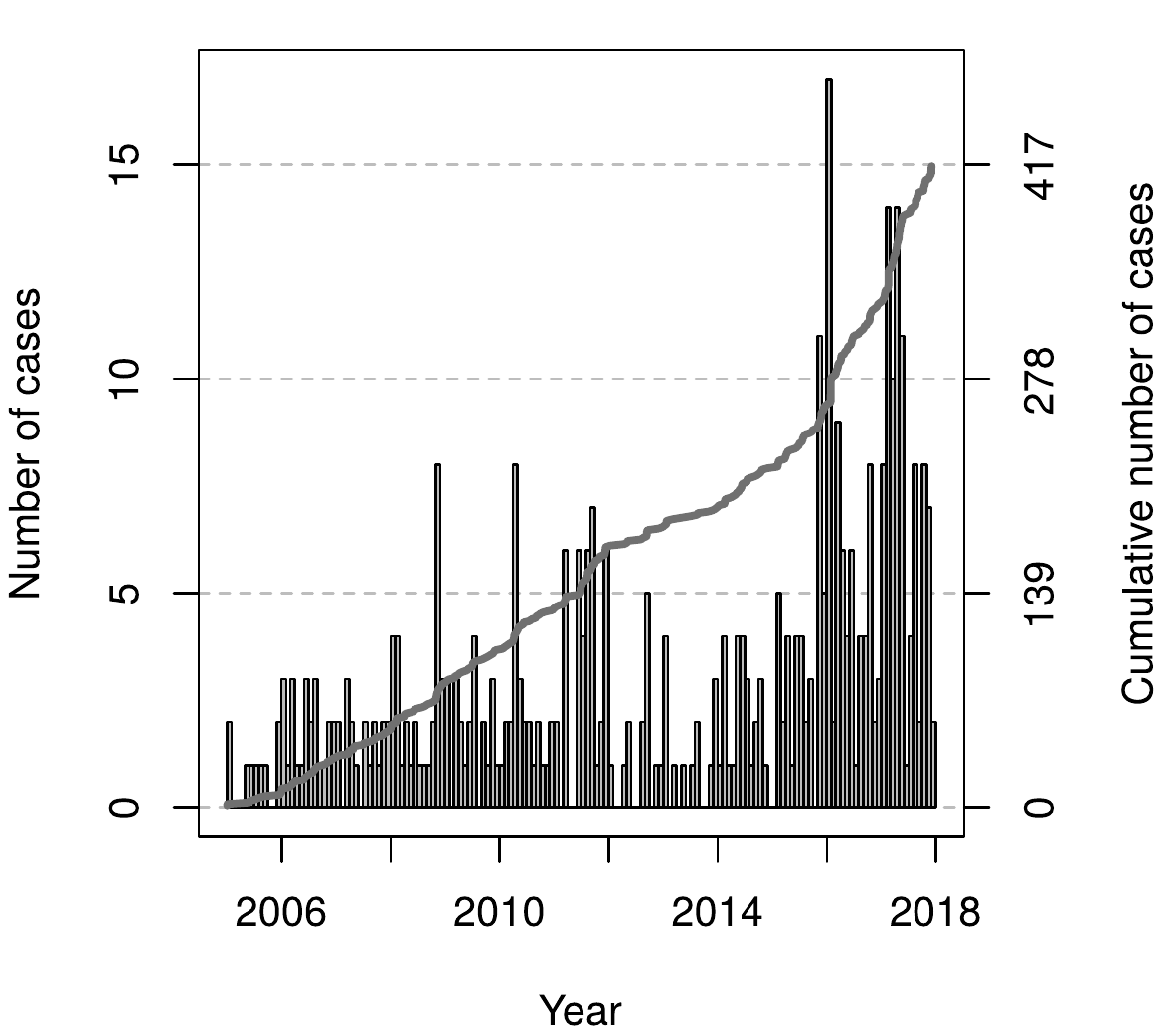}
\caption{The number of far-right extremists exposed in the PIRUS database, both per-year (left) and cumulative (right), between 2005 and 2017.}

\label{cases}
\end{figure}

\begin{figure*}[t]
\centering
\includegraphics[width=\linewidth]{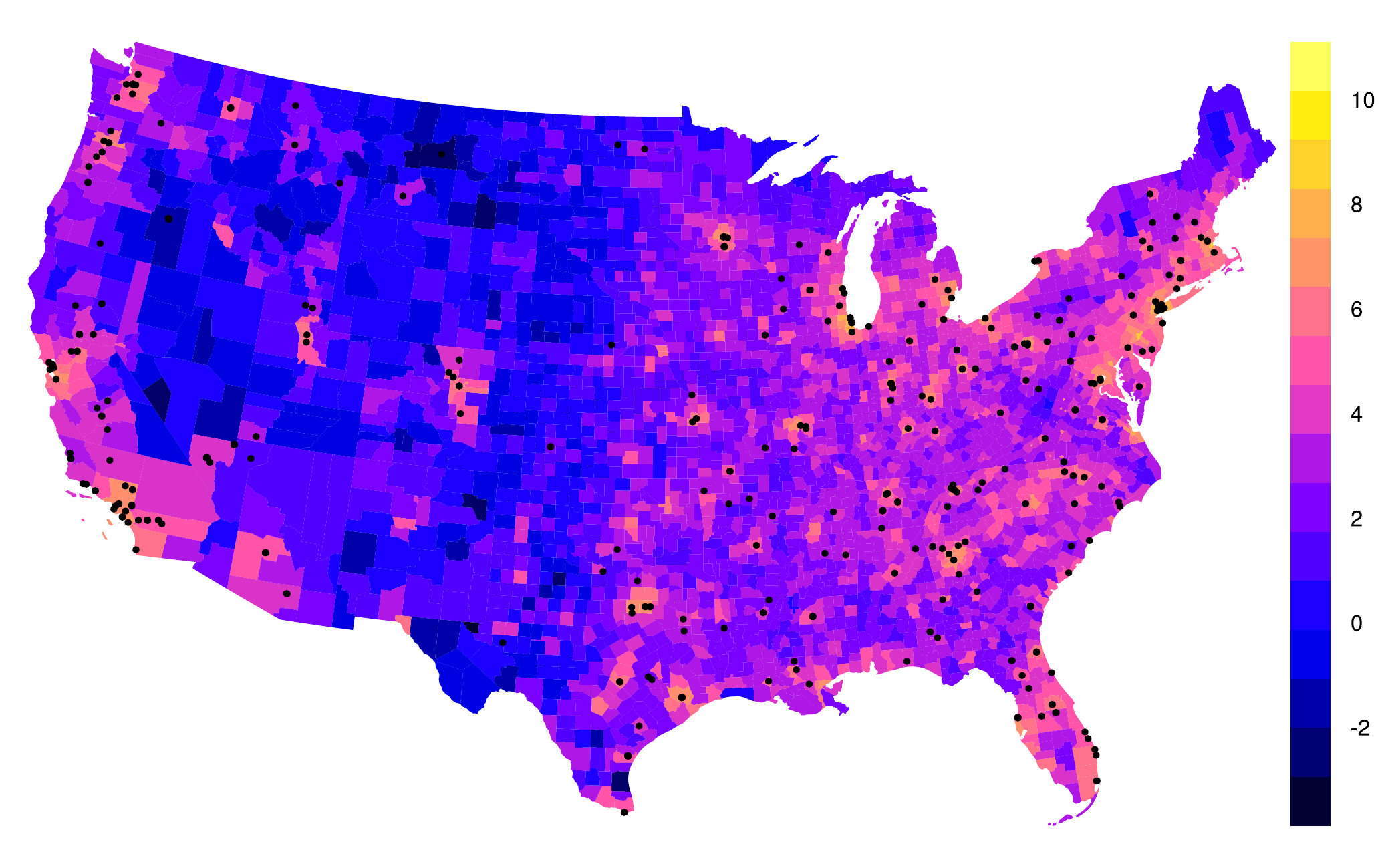}
\caption{The untied locations of far-right extremists exposed in the PIRUS database between 2005 and 2017. The color of each county corresponds to its log-transformed population density.}
\label{map}
\end{figure*}

All individual-level data came from PIRUS. Only individuals with far-right ideology who were exposed during or after 2005 (the earliest year with social media data) with location data at the city-level or lower (\textit{n} = 416; F: 6.0\%, M: 94.0\%) were included (see Figures \ref{cases} and \ref{map}). For each individual, the date and location of their exposure (usually when their activity/plot occurred), whether their plot was successful (34.9\%), the anticipated fatalities of their plot (0: 69.5\%, 1-20: 26.0\%, $>$20: 2.6\%, $>$100: 1.9\%), whether they were a member of a formal or informal group of extremists (58.4\%), and whether social media played a role in their radicalization (31.2\%), were included. Unknown or missing values for each predictor (plot success: 0.5\%, anticipated fatalities: 13.5\%, group membership: 0\%, social media: 54.8\%) were coded as 0. To ensure that the coding procedure for missing predictor values did not introduce bias, I checked whether the results of the full model were consistent after multiple imputation with chained equations and random forest machine learning (see Table \ref{impute_check}). The location of each exposure was geocoded from the nearest city or town using the R package \textit{ggmap} \cite{Kahle2013}. Since domestic terrorists tend to commit acts in their local area \cite{Smith2008,Klein2017,Marchment2018}, I assumed that exposure locations reflect where individuals were radicalized.

State-level gun ownership was estimated using a proxy measure based on suicide rates and hunting licenses \cite{Seigel2014}. Using data from 2001, 2002, and 2004 (the only three years for which state-level gun ownership data is available), Seigel et al. found that the following proxy correlates with gun ownership with an $R^2$ of 0.95:

\begin{equation}
\label{gun_proxy}
(0.62 \cdot \frac{FS}{S})+(0.88\cdot HL)-0.0448
\end{equation}

\noindent where $FS/S$ is the proportion of suicides that involve firearms (from the Centers for Disease Control and Prevention\footnote{\url{https://www.cdc.gov/injury/wisqars/index.html}}, or CDC), and $HL$ is hunting licenses per capita (from the United States Fish and Wildlife Service\footnote{\url{https://wsfrprograms.fws.gov/Subpages/LicenseInfo/Hunting.htm}}) \cite{Seigel2014}. Missing suicide rates (five years for DC, two years for Rhode Island) were replaced with the mean values for that state. State-level hate group data was collected from the Southern Poverty Law Center\footnote{\url{https://www.splcenter.org/hate-map}}, while violent crime data was collected from the Federal Bureau of Investigation's Uniform Crime Reporting Program\footnote{\url{https://www.fbi.gov/services/cjis/ucr}}.

County-level demographic data was collected from the US Census using the R package \textit{censusapi} \cite{Recht2019}. This included population density, poverty rate, Gini index of income inequality, percentage of the population that is non-white, percentage of the population that has at least a high-school diploma, and unemployment rate. County-level income, race, education, and unemployment data is only available after 2009, so the 2010 data was used for 2005-2009. County-level presidential election voting records were collected from the Massachusetts Institute of Technology Election Lab\footnote{\url{https://electionlab.mit.edu/data}}, and non-election years were assigned the data from the most recent election year.

Geographic data was collected from the US Census using the R package \textit{tigris} \cite{Walker2019}.

\subsection{Model Specification}

Twinstim modeling was conducted using the R package \textit{surveillance} \cite{Meyer2017}. To convert the data to a continuous spatio-temporal point process, all tied locations and dates were shifted in a random direction up to half of the minimum spatial and temporal distance between events (1.52 km and 0.5 days, respectively) \cite{Meyer2014}.

Step functions were used to model both spatial and temporal interactions. Visual inspection of the pair correlation function for the point pattern indicates that the data is significantly clustered up to 400 km (see Figure \ref{pcf}). As such, the spatial step function was split into four 100 km intervals with 400 km as the maximum interaction radius \cite{Nightingale2015}. The temporal step function was split into four six-month intervals up to two years (based on the the high degree of variation in radicalization and attack planning times among domestic extremists \cite{Smith2009,Silkoset2016,Bouhana2018}). I attempted the analysis with different combinations of power-law, Gaussian, and Student spatial functions, and exponential temporal functions, but these variations converged to unrealistically steep spatial and temporal interaction functions that approached zero around two km and two days, and appeared to be significantly influenced by the tie-breaking procedure \cite{Meyer2014}.

Population density (county-level) was log-transformed and used as an offset endemic term. A centered time trend was also included to determine whether the strength of the endemic component has shifted over time. Poverty rate (county-level), Gini index of income inequality (county-level), gun ownership (state-level), percentage of the population that is non-white (county-level), percentage of the population that has at least a high-school diploma (county-level), unemployment rate (county-level), percentage of voters that vote Republican in presidential elections (county-level), violent crime rate per thousand residents (state-level), and number of hate groups per million residents (state-level) were included as dynamic endemic predictors that change annually. Plot success, anticipated fatalities, group membership, and social media radicalization were included as epidemic predictors.

All possible models with all possible combinations of predictors were run and ranked by Akaike’s Information Criterion (AIC) \cite{Burnham2002}. The best fitting model with the lowest AIC was used assess the effects of each variable on event probability. Rate ratios were calculated by applying exponential transformation to the model estimates.

\subsection{Permutation Test}

To determine whether the spatio-temporal interaction of the epidemic component was statistically significant, I used the Monte Carlo permutation approach developed by Meyer et al. \cite{Meyer2016}. Using this approach, a twinstim model with all endemic predictors from the best fitting model and no epidemic predictors was compared to 1,000 permuted null models with randomly shuffled event times. For each permutation I estimated the reproduction number (\textit{R\textsubscript{0}}), or the expected number of future events that an event triggers on average, which represents ``infectivity''. A \textit{p}-value was calculated by comparing the observed \textit{R\textsubscript{0}} with the null distribution of the subset of permutations that converged.

For additional support, I also ran a likelihood ratio test and a standard Knox test of spatio-temporal clustering. The Knox test was conducted with spatial and temporal radii of 100 km and six months (the upper bounds of the first steps in the step functions), respectively \cite{Knox1964}.

\subsection{Simulations}

To further assess the quality of the model, I conducted simulations from the cumulative intensity function using Ogata's modified thinning algorithm according to Meyer et al. \cite{Meyer2012}. Using the parameters of the best fitting model, I conducted 1,000 simulations of the last six months of the study period and compared the results to the observed data.

\section*{Results}

\begin{figure}[t]
\centering
\includegraphics[width=0.9\linewidth]{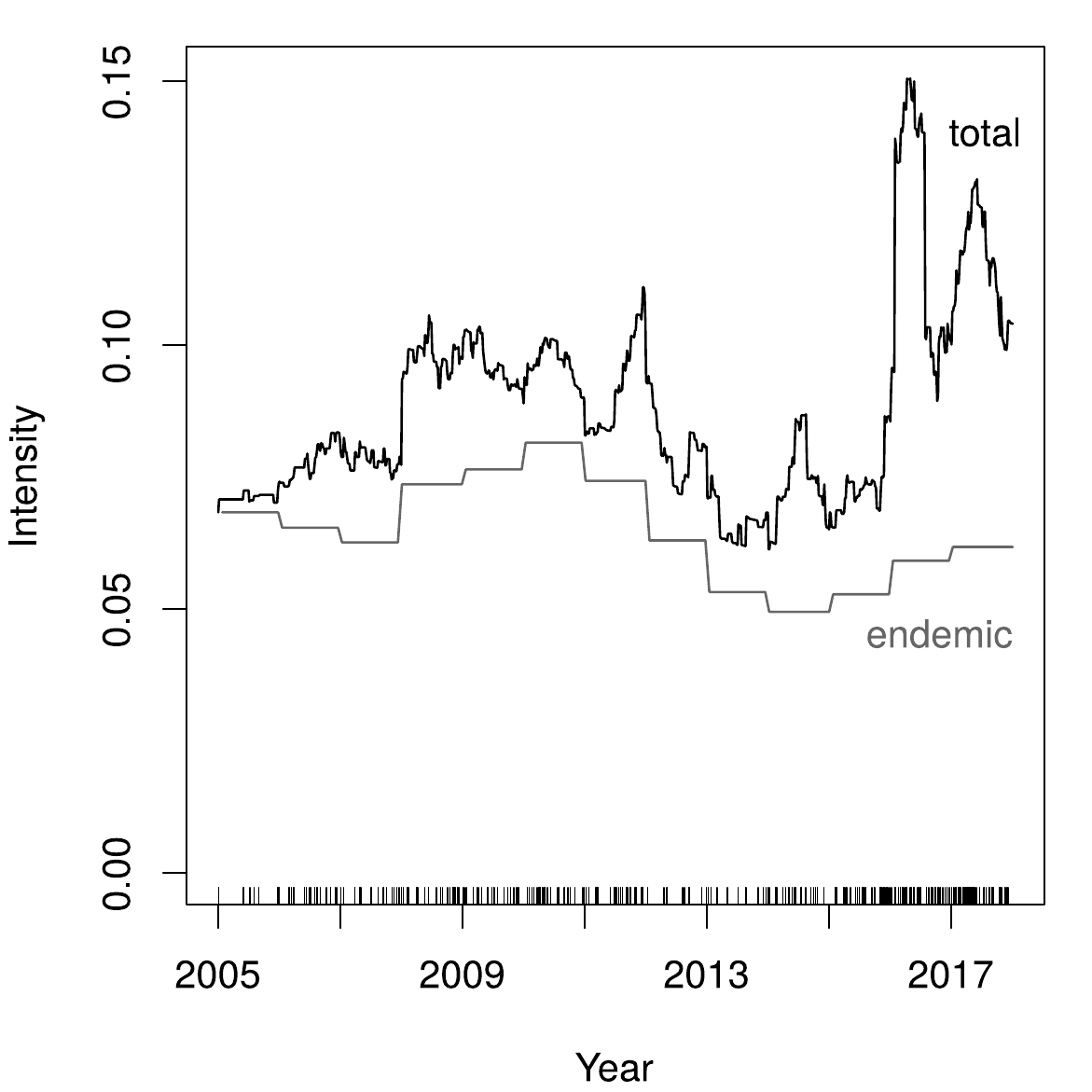}
\caption{The total intensity (in black), as well as the isolated endemic component (grey), over time. Total intensity can be interpreted as the proportion of radicalization probability that is explained by the endemic and epidemic components.}
\label{intensity}
\end{figure}

The results of the best fitting model ($\Delta$AIC $<$ 2), which included seven endemic and two epidemic predictor variables, are shown in Table \ref{twinstim_output}. Firstly, there is a statistically significant time trend whereby the endemic rate decreases by 4.6\% each year, indicating that the strength of the epidemic component has increased over time. There appears to be a baseline increase in the endemic component between 2008-2012 which likely corresponds to the financial crisis \cite{Funke2016}, as well as a significant spike in the epidemic component around 2016 which likely corresponds to the presidential election \cite{Rushin2018,Giani2019} (Figure \ref{intensity}). There are also significant positive effects of poverty rates (\textit{p} $<$ 0.01) and the presence of hate groups (\textit{p} $<$ 0.0001) on radicalization probability. Interestingly, the percentage of voters that vote Republican in presidential elections (\textit{p} $<$ 0.0001), the percentage of the population that is non-white (\textit{p} $<$ 0.05), and unemployment rates (\textit{p} $<$ 0.0001) appear to have significant negative effects on radicalization probability. Gun ownership, education level, and violent crime all had no significant effect on radicalization probability. When Republican voting was replaced with the absolute percent difference between Republican and Democratic voting, a proxy measure for the competitiveness of elections, it was no longer significant. A variance inflation factor test identified no collinearity problems among the time-averaged endemic predictors (\textit{VIF} $<$ 3) \cite{Zuur2010}.

\begin{table}[b]
\centering
\begin{tabular}{lrrr}
\hline
& RR & 95\% CI & \textit{p}-value \\ 
\hline
Time trend & 0.946 & 0.91--0.98 & 0.0015 \\
Poverty & 1.052 & 1.02--1.09 & 0.0025 \\
Unemployment & 0.871 & 0.82--0.93 & $<$0.0001 \\
Republican voting & 0.969 & 0.96--0.98 & $<$0.0001 \\
Non-white population & 0.989 & 0.98--1.00 & 0.038 \\
Education level & 0.980 & 0.96--1.00 & 0.075 \\
Hate groups & 1.191 & 1.12--1.27 & $<$0.0001 \\
Violent crime & 0.930 & 0.84--1.03 & 0.17 \\
\hline
Group membership & 4.563 & 1.57--13.28 & 0.0054 \\
Social media & 2.722 & 1.55--4.76 & 0.0005 \\
\hline
\end{tabular}
\caption{The results of the twinstim modeling, with estimated rate ratios (RR), Wald confidence intervals, and \textit{p}-values. Endemic predictors (including the overall time trend) are in the top portion of the table, whereas epidemic predictors are in the bottom.}
\label{twinstim_output}
\end{table}

\begin{figure}[t]
\centering
\includegraphics[width=0.9\linewidth]{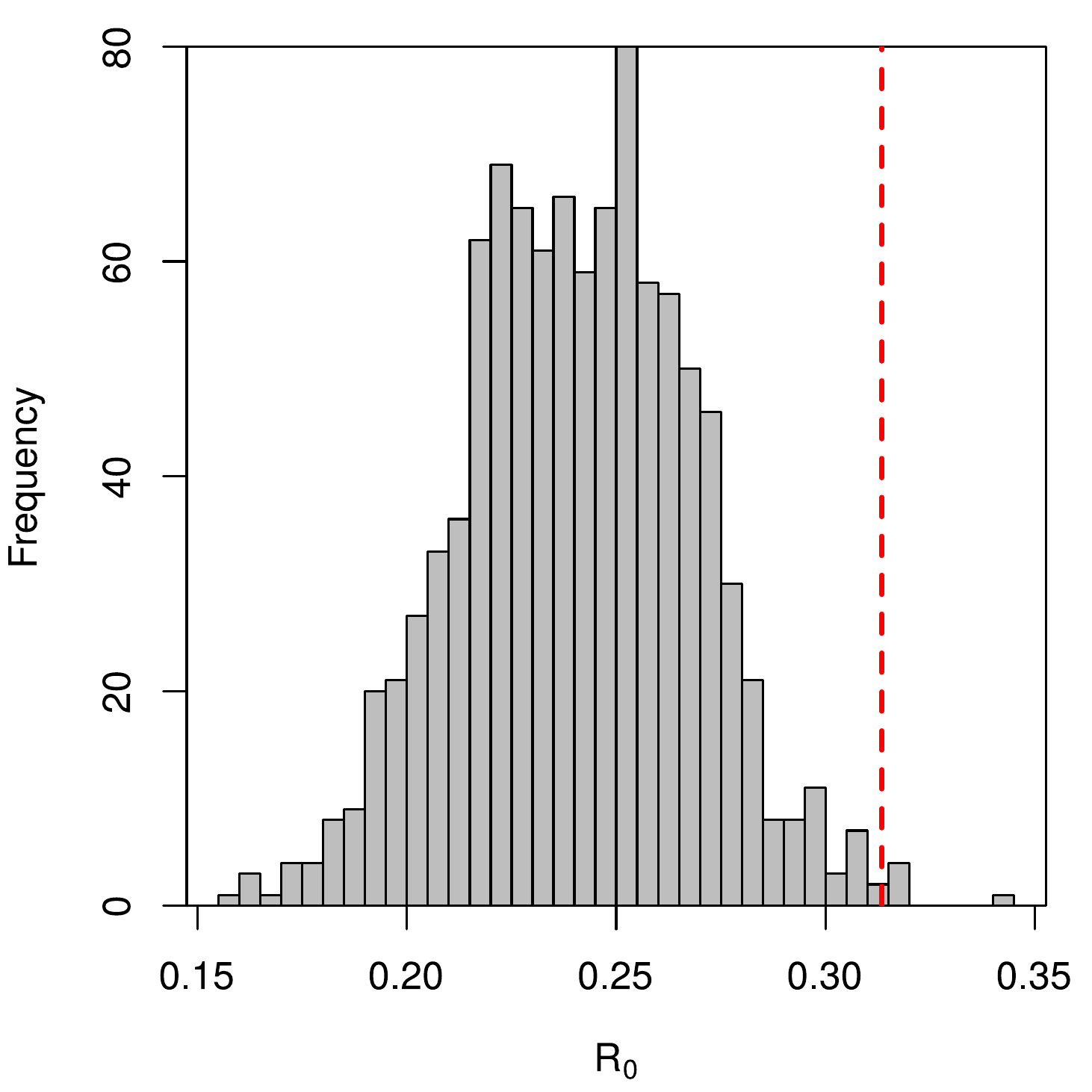}
\caption{The results of the Monte Carlo permutation test. The grey bars show the null distribution of \textit{R\textsubscript{0}} from the 739 permutations that converged, whereas the red dashed line shows the observed \textit{R\textsubscript{0}} (0.31) calculated from the twinstim model.}
\label{epitest}
\end{figure}

Both group membership and radicalization via social media have strong and significant positive effects on epidemic probability. Exposures of individuals who belong to formal or informal extremist groups are over four times more likely to be followed by future exposures in close spatial or temporal proximity (\textit{p} $<$ 0.01). Similarly, exposures of individuals radicalized on social media are almost three times as likely to be followed by future exposures (\textit{p} $<$ 0.01). Anticipated fatalities and plot success did not appear in the best fitting model. Estimates of the decaying spatial and temporal interaction functions, as well as model diagnostics, can be seen in Figures \ref{siaf_tiaf} and \ref{residuals}, respectively. A variance inflation factor test identified no collinearity problems among the epidemic predictors (\textit{VIF} $<$ 3) \cite{Zuur2010}.

Based on the permutation test, the observed \textit{R\textsubscript{0}} (0.31) is significantly higher than the null distribution of the converged permutations (\textit{N\textsubscript{conv}} = 739, \textit{p} $<$ 0.01) (Figure \ref{epitest}). This indicates that the spatio-temporal interaction in the epidemic model is significant. Both the likelihood ratio test of the epidemic against the endemic model (\textit{p} $<$ 0.0001) and the Knox test (\textit{p} $<$ 0.0001) support this result.

\begin{figure}
\centering
\includegraphics[width=0.9\linewidth]{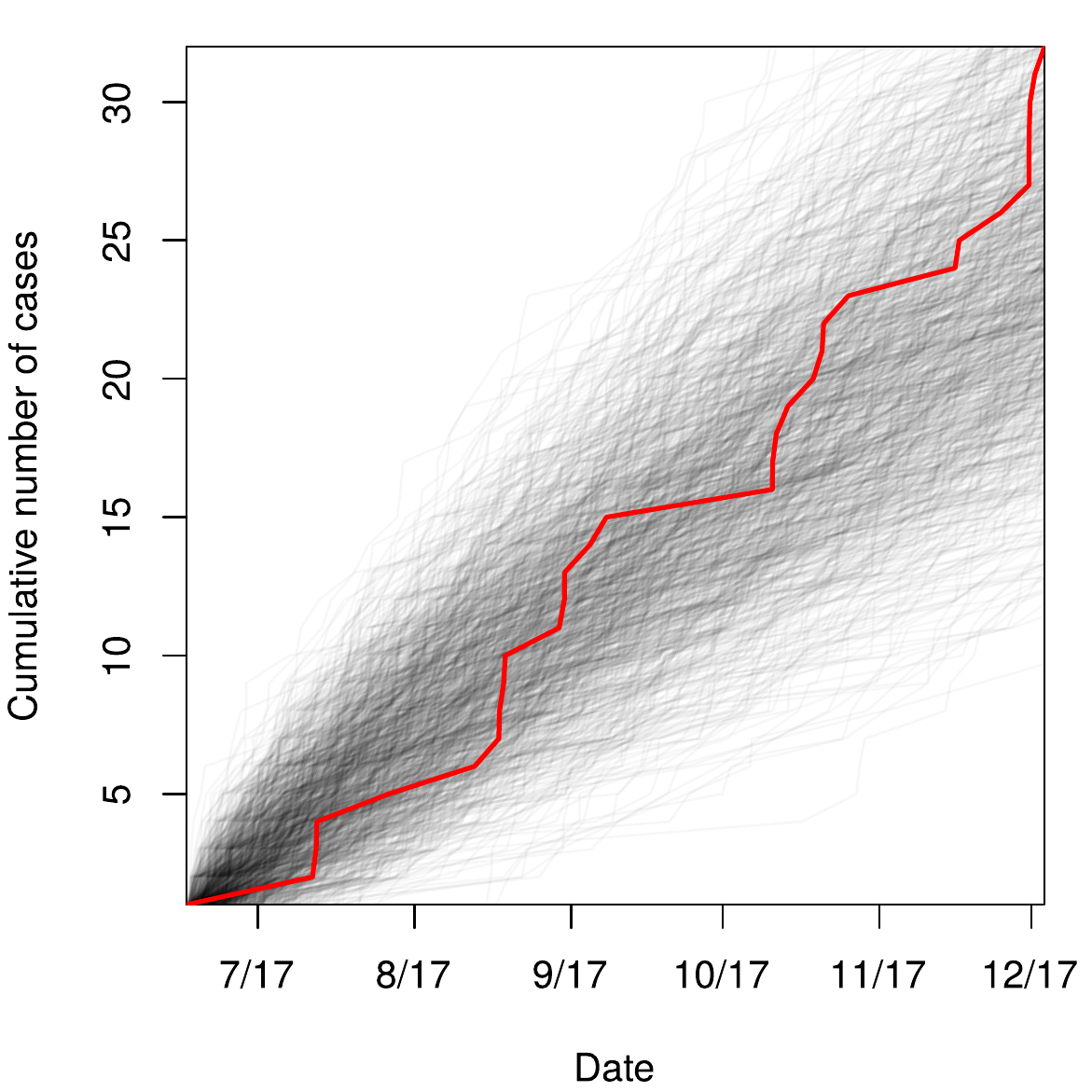}
\caption{The cumulative number of exposures during the last six months of the study period (red), as well as the results of 1,000 simulations (grey).}
\label{pred_time}
\end{figure}

\begin{figure}
\centering
\includegraphics[width=1\linewidth]{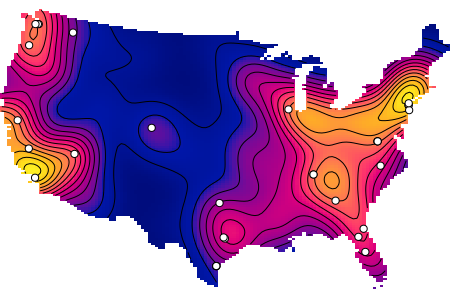}
\caption{The exposure locations during the last six months of the study period. The color gradient, ranging from blue (low) to yellow (high), represents a Gaussian kernel density for the results of 1,000 simulations with a bandwidth of 200 km. The contour lines segment the kernel density into 10 levels.}
\label{pred_space}
\end{figure}

The results of the simulations can be seen in Figures \ref{pred_time} and \ref{pred_space}. On average the simulations neatly match the observed cumulative number of exposures between June, 2017 and January, 2018 (Figure \ref{pred_time}), indicating that the model accurately captures the temporal dynamics in the data. Similarly, the model appears to do a good job of capturing the spatial dynamics in the data, although it is clearly weighted towards high population density areas (Figure \ref{pred_space}).

\section*{Discussion}

By applying novel epidemiological methods to data on 416 extremists exposed between 2005 and 2017, this study provides evidence that patterns of far-right radicalization in the United States are consistent with a contagion process. Firstly, the estimated reproduction number is significantly higher than those from simulated null models, indicating that endemic causes alone are not sufficient to explain the spatio-temporal clustering observed in the data. The reproduction number for radicalization (\textit{R\textsubscript{0}} = 0.31) is also lower than one, suggesting that extremist ideologies behave like complex contagions that require reinforcement for transmission. Fortunately, this means that extremist ideologies are unlikely to spread uncontrollably through populations like seasonal influenza (\textit{R\textsubscript{0}} = 1.28) \cite{Biggerstaff2014}, but outbreaks can occur under the right endemic and epidemic conditions. For example, regions with higher rates of poverty and hate group activity are more likely to experience far-right extremism, whereas regions with a larger non-white population, more Republican voting, and higher rates of unemployment are less likely to experience far-right extremism. Most importantly, radicalizations involving extremist groups or social media significantly increase the epidemic probability of future radicalizations in the same location. This suggests that clusters of radicalizations in space and time are driven by activism and organizing rather than a copy-cat effect.

The fact that group membership significantly increases the epidemic strength of events, and the presence of hate groups significantly increases radicalization probability, suggests that local organizing remains a potent recruitment tool of the far-right movement. This idea is reflected in recent increases in rallies across the country, such as ``Unite the Right'' in Charlottesville, VA in August of 2017, that have been attended by regional chapters of white nationalist and militia organizations. It also suggests that concerns about typological ``lone wolves'' radicalized over social media should not overshadow the persistent and expanding far-right movement in the United States. Only 10.8\% of people in this study were radicalized on social media independently of an extremist group, indicating that solo actors are still the minority in the far-right movement. That being said, solo actors radicalized on social media, such as Omar Mateen (Pulse nightclub shooting in 2016) and Dylann Roof (Charleston church shooting in 2015) \cite{Holt2019}, are typically deadlier than group members in the United States \cite{Phillips2017}, and should thus be the subject of much future research.

Radicalization on social media also significantly increases the epidemic strength of events, indicating that social media platforms augment physical organizing and that the diffusion of extremist ideologies online is likely geographically biased. The increasing role of social media in far-right extremism and radicalization is well established \cite{Holt2019,Ottoni2018,Costello2018,Lowe2019,Winter2019}. Social media platforms like Twitter provide extremist communities with low cost access to large audiences that might not otherwise engage with far-right content \cite{Wu2015,Bertram2016}. For example, one report found that only 44\% of people who follow high-profile white nationalists on Twitter overtly express similar ideologies \cite{Berger2013}. As mainstream platforms clamp down on hate speech, extremist users have just shifted their traffic to alternative sites such as 8chan and Gab \cite{Blackbourn2019,Hodge2019}. Given the centrality of social media in far-right organizing, future research should explore how counter-narratives \cite{VanEerten2017,Voogt2017} and other strategies could be used to fight the spread of extremist ideology online.

The results indicate that county-level poverty rates increase the probability of far-right radicalization. While there is little to no evidence that poverty predicts extremism at the state-level \cite{Piazza2017,Gale2002,Lin2018,Durso2013}, studies at the county-level have found that poverty predicts both mass shooting rate \cite{Kwon2019a} and hate groups (presence \cite{Medina2018} not longevity \cite{Suttmoeller2015,Suttmoeller2016,Suttmoeller2018}). This discrepancy between geographic resolutions indicates that using state-level poverty data obscures local variation. The results of this study also reveal a negative effect of unemployment rate on radicalization, adding to the remarkably contradictory evidence for links between unemployment and extremism in the United States \cite{Jefferson1999,Goetz2012,Piazza2017,Majumder2017,Green1998,Gale2002,Espiritu2004}. Although this result appears to be counter-intuitive, I hypothesize that poverty and unemployment may interact in driving radicalization. For example, individuals from regions where jobs are plentiful but poverty remains high may be the most disillusioned and susceptible to extremist ideologies. Interestingly, income inequality did not appear in the best fitting model, and had no significant effect when included. This suggests that overall deprivation, as measured by poverty rates, is more important in driving radicalization than inequality. Previous studies that have found a positive impact of income inequality on hate groups or crimes either used state-level data \cite{Majumder2017}, did not account for poverty rate \cite{McVeigh2004,Goetz2012}, or combined it with poverty rate into a single index \cite{McVeigh2012}. Interestingly, both unemployment \cite{Pah2017} and income inequality \cite{Kwon2017,Kwon2019a,Kwon2019b} appear to be strong predictors of mass shootings. Although this seems paradoxical, the majority of mass shootings are not ideologically driven \cite{Capellan2015}, so the socioeconomic drivers may be different than for far-right radicalization.

Violent crime appears to have no influence on radicalization. Although one study of the Klu Klux Klan found that high levels of far-right activity can increase homicide rates in the long-term \cite{McVeigh2012}, there is little evidence for violent crime rates driving increases in extremist violence or radicalization \cite{Sweeney2018}. Hate crime is only very weakly correlated with violent crime \cite{Gladfelter2017}, and extremist violence is even more rare \cite{LaFree2009}, so they are likely driven by different factors.

Previous studies have found strong evidence for a negative relationship between education and hate crime rates \cite{Espiritu2004,Gladfelter2017}, a positive relationship between education and mass shooting rates \cite{Kwon2017,Kwon2019b}, and no evidence for a relationship between education and hate group organizing \cite{Durso2013,McVeigh2014,Florida2011}. The results of this study are consistent with the latter category, which makes sense given that the majority of the plots in the dataset were non-violent.

The negative effect of Republican voting on event probability could be because individuals on the far-right of the political spectrum who live in counties with more Democratic voters may feel more partisan hostility \cite{Miller2015}. Interestingly, this effect does not appear to be the result of more competitive elections \cite{Suttmoeller2015}, as the absolute difference between Republican and Democratic voting did not significantly influence event probability. Alternatively, the negative effect of Republican voting may be due to the fact that many of the rural counties that lean heavily Republican have low population densities and no recent history of extremist violence. A previous study that found mixed evidence for a positive influence of Republican voting on the presence of hate groups excluded counties without hate groups from the modeling, which may have eliminated this skew effect \cite{Medina2018}.

The fact that the percentage of the population that is non-white negatively predicts far-right extremist violence is consistent with the intergroup contact hypothesis, which suggests that prolonged contact between racial groups reduces conflict under certain conditions \cite{Allport1954}. Although other researchers have suggested that population heterogeneity increases far-right radicalization \cite{McVeigh2004}, the only study to find evidence of this in the United States did not explicitly account for population density \cite{LaFree2014}. Other studies controlling for population density have found that both anti-black hate crimes and hate groups appear to be more common in white dominated, racially homogeneous areas \cite{Gladfelter2017,Medina2018}. Despite mixed evidence for the intergroup contact hypothesis, it is widely accepted that community diversity and tolerance is key to fighting radicalization and extremist violence globally \cite{Gunaratna2013,Ellis2017,Ercan2017,Hoffman2018,UnitedNationsDevelopmentProgramme2016,SouthernPovertyLawCenter2017}.

Gun ownership does not predict radicalization in this model, which is unsurprising since only 30.5\% of people in this study planned on committing fatal attacks but interesting given the centrality of gun control in debates following mass shootings in the United States \cite{Pierre2019,Joslyn2017,Luca2020}. Despite strong evidence that gun ownership is linked to mass shooting rates at the national-level \cite{Reeping2019}, evidence for same pattern at the state-level remains mixed. Previous studies have found that it either positively predicts mass shootings overall \cite{Reeping2019}, when combined with particular gun control laws \cite{Anisin2018}, or not at all \cite{Lin2018,Pah2017}. Unfortunately CDC funding for research on gun ownership was restricted by Congress in 1996 after lobbying by the National Rifle Association, so potential links between extremist violence and gun ownership remain understudied \cite{Lemieux2014,Winker2016,Morrall2018,DeFoster2018}.

Several limitations of this study should be highlighted. Firstly, the PIRUS database only represents a subset of radicalized individuals in the United States. The creators of the database used random sampling to maximize its representativeness over different time periods, but there remains a possibility of spatial or temporal bias in the original data due to factors like law enforcement effort. In addition, the geographic locations of events are only geocoded to the city-level, potentially enhancing the spatial clustering of the data. Furthermore, social media data were missing for a significant number of individuals (54.8\%). The significance level of the estimate for social media usage is extremely low and robust to imputation, indicating that it likely reflects a real effect, but researchers should exercise caution when interpreting this result \cite{Safer-Lichtenstein2017}. Lastly, the spatial resolution of three of the endemic predictors was limited to the state-level, which may have flattened some important local variation. One of these variables, gun ownership, was also a proxy measure. Policymakers should release historical restrictions on research funding for gun violence and hate crime research to improve data resolution for future studies.

In conclusion, far-right radicalization in the United States appears to spread through populations like a complex contagion. Both social media usage and group membership enhance the contagion process, indicating that online and physical organizing remain primary recruitment tools of the far-right movement. In addition, far-right radicalization is more likely in Democratic-majority regions with high poverty and low unemployment, fewer non-white people, and more hate group activity. While the federal government has acknowledged the threat of far-right extremism \cite{TheDepartmentofHomelandSecurity2019}, funding for organizations researching or fighting the movement has decreased in recent years \cite{OToole2019}. Based on the results of this study, I recommend that policymakers reconsider their funding priorities to address the expanding far-right extremist movement in the United States. Future research should investigate how specific interventions, such as online counter-narratives to battle propaganda, may be effectively implemented to mitigate the spread of extremist ideology.

\section*{\large Data \& Code Availability Statement}

All data used in the study are available online either publicly or upon request (PIRUS). The R scripts used in the study will be made available upon peer-reviewed publication.

\renewcommand*{\bibfont}{\scriptsize}
\printbibliography[title=\large References]

\end{refsection}

\newpage
\onecolumn

\begin{refsection}

\begin{center}
	\sffamily\LARGE Supporting information\rmfamily
\end{center}

\setcounter{figure}{0}
\setcounter{table}{0}
\setcounter{subsection}{0}
\renewcommand{\thetable}{S\arabic{table}}
\renewcommand{\thefigure}{S\arabic{figure}}

\subsection{Imputation Check}

To ensure that the coding procedure did not bias the estimation of the epidemic predictors, the full twinstim model was re-run after multiple imputation with chained equations using the R package \textit{mice} \cite{vanBuuren2011} and random forest machine learning using the R package \textit{missForest} \cite{Stekhoven2012}. All four epidemic predictors (plot success, anticipated fatalities, and group membership) were used in fitting and training. The maximum iterations was set to 10 and number of trees was set to 100. The results of 100 rounds of both imputation methods can be seen in Table \ref{impute_check}.

\begin{table}[ht]
\centering
\begin{tabular}{lrrrr}
\hline
& \multicolumn{2}{c}{Chained equations} & \multicolumn{2}{c}{Random forest} \\
\hline
& RR & \textit{p}-value & RR & \textit{p}-value \\ 
\hline
Group membership & 5.014 & 0.0040 & 6.61 & 0.00040 \\
Social media & 2.29 & 0.065 & 4.092 & 0.00013 \\
Anticipated fatalities & 1.15 & 0.43 & 0.90 & 0.53 \\
Plot success & 0.93 & 0.78 & 1.11 & 0.55 \\
\hline
\end{tabular}
\caption{The average rate ratios and \textit{p}-values for all epidemic predictors after 100 rounds of imputation and estimation using the full model.}
\label{impute_check}
\end{table}

Since the observed estimate of social media, the only epidemic predictor with missing data in the best fitting model, is between those from the two imputation methods, and as random forest in \textit{missForest} outcompetes chained equations in \textit{mice} in most \cite{Stekhoven2012,Waljee2013,Liao2014,Muharemi2018,Misztal2019,Cui2019} (but not all \cite{Shah2014,Penone2014}) direct comparisons, I assume that the coding method did not significantly influence the results. I assumed that data were missing at random, although missing social 

\subsection{Spatial interaction}

\begin{figure}[h]
\centering
\includegraphics[width=0.7\linewidth]{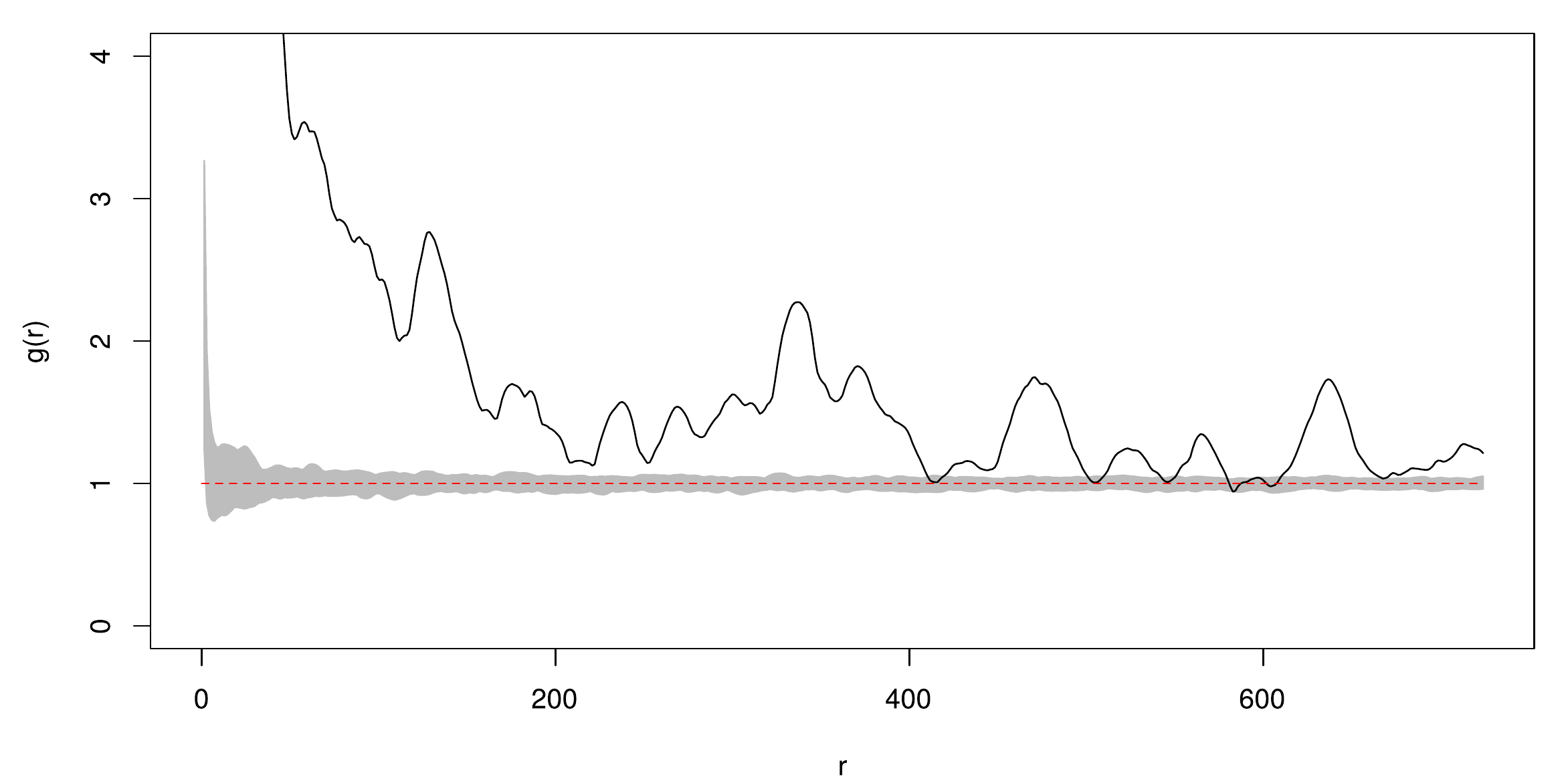}
\caption{The pair correlation function at different pairwise distances in km (\textit{x}-axis). The black line is the observed function for the data, the red line is the theoretical function assuming spatial randomness, and the grey envelope shows the upper and lower bounds of the functions from 100 simulated point patterns demonstrating spatial randomness.}
\label{pcf}
\end{figure}

\begin{figure}[h]
\centering
\includegraphics[width=0.8\linewidth]{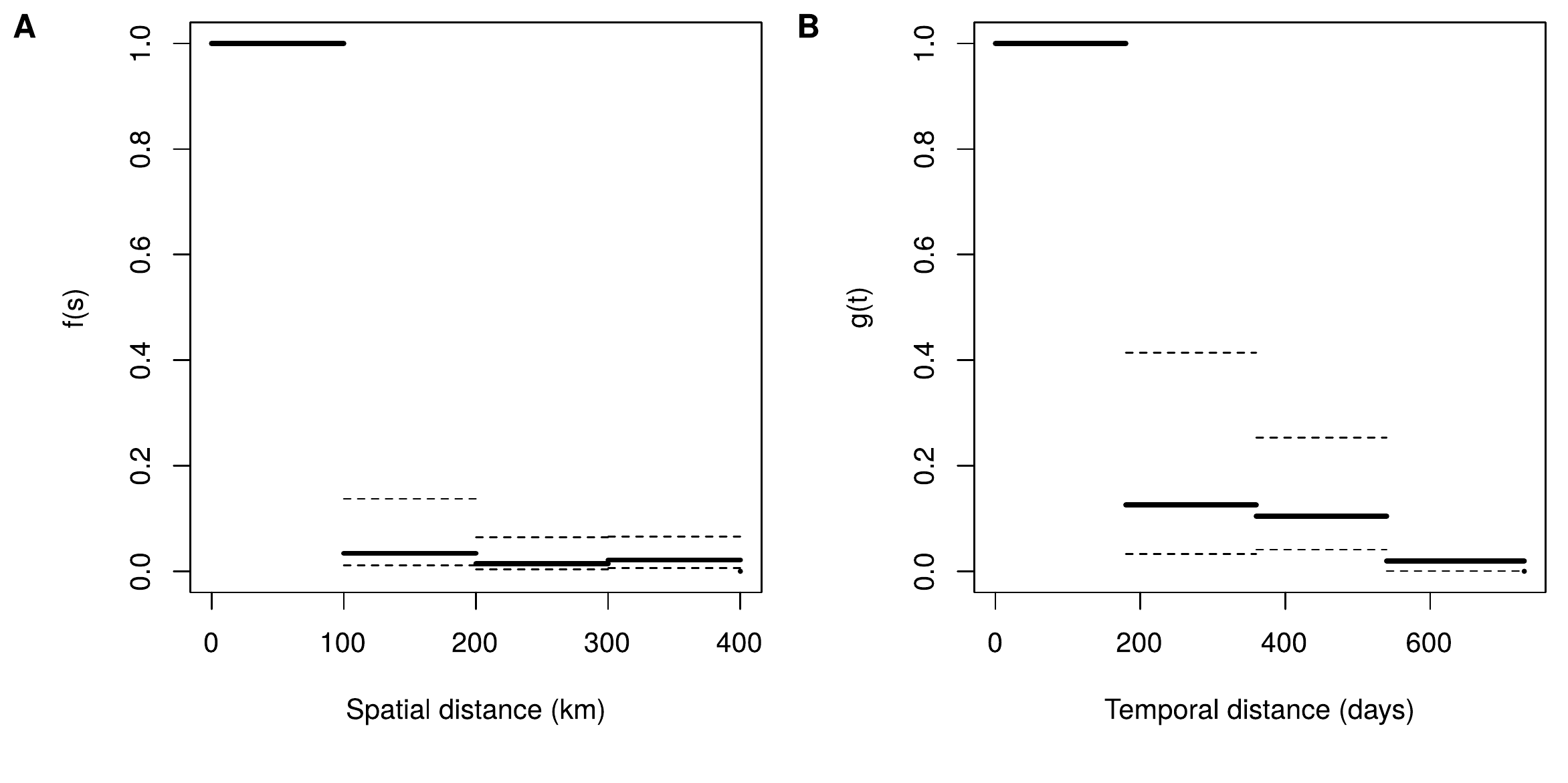}
\caption{Estimates of the scaled spatial (left panel) and temporal (right) step functions. The 95\% Monte Carlo confidence intervals were each calculated from 100 samples.}
\label{siaf_tiaf}
\end{figure}

\subsection{Diagnostics}

\begin{figure}[h]
\centering
\includegraphics[width=0.8\linewidth]{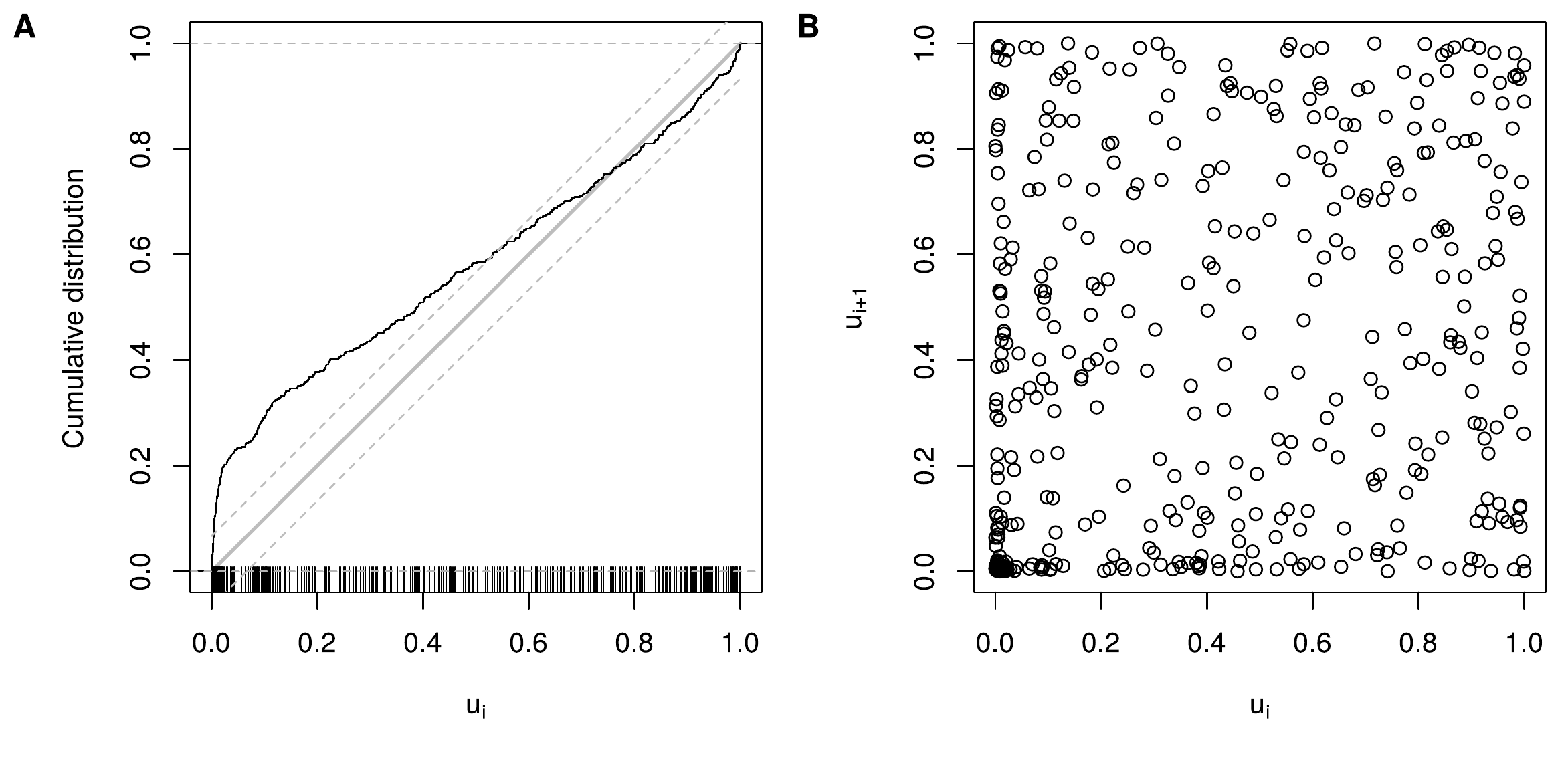}
\caption{(A) The empirical cumulative density function of U\textsubscript{i}, or the standardized residuals according to Ogata \cite{Ogata1988}, with 95\% Kolmogorov-Smirnov confidence bands. (B) A scatterplot of U\textsubscript{i} and U\textsubscript{i+1} to look for serial correlation.}
\label{residuals}
\end{figure}

The residuals, or the fitted cumulative intensities over time, were calculated and transformed to fit a uniform distribution according to Ogata \cite{Ogata1988}. The cumulative density function diverges from expectations for U\textsubscript{i} $<$ 0.58, which appears to be the result of tie-breaking with small temporal distances (0.5 days) \cite{Meyer2012}. Increasing the tie-breaking distance to $>$ 20 days to improve the cumulative density function and reduce serial correlation did not significantly change the predictor estimates, so I chose to use the original model.

\clearpage

\renewcommand*{\bibfont}{\scriptsize}
\printbibliography[title=\large References]

\end{refsection}

\end{document}